\documentclass{article}
\usepackage[utf8]{inputenc}
\usepackage{authblk}
\usepackage{graphicx}
\usepackage{siunitx}
\title{Electronic readout of optically excited surface plasmons}

\author[a]{Alec R. Cheney}
\author[a]{Borui Chen}
\author[b]{Tim Thomay\footnote{Corresponding author:T.T.: E-mail: thomay@buffalo.edu}}
\affil[a]{Department of Electrical Engineering, University at Buffalo, State University of New York, Buffalo, New York 14260, USA}
\affil[b]{Department of Physics, University at Buffalo, State University of New York, Buffalo, New York 14260, USA
}
\date{}
\begin{document}
\maketitle
\begin{abstract}
Leveraging thermal losses as a useful consequence of surface plasmons in metal nanostructures has gained traction in recent years. This thermalization of hot electrons also induces a resistance change to an applied bias current, which we use to realize an all electronic readout of surface plasmons. The interplay of the plasmonic k-vector dependence and the applied bias current allows us to distinguish between linear polarizations of an incident laser beam for polarimetry and polarization imaging uses. This illustrates the potential applications this technique offers as a fully CMOS compatible plasmonic sensor. Moreover, we demonstrate an electronic signal that depends on the delay between two laser pulses on ultrafast timescales, providing insight into the highly non-equilibrium dynamics of the hot electron distribution inside the metal. Using an electronic approach to surface plasmons broadens access and simplifies existing applications, while simultaneously opening the door to new pathways for developing integrated sensors for processes on ultrafast timescales. \newline
\end{abstract}

The interaction of light with free electrons in metal nanostructures and the associated strong local electric field enhancement has led to exciting developments within the field of plasmonics.
For example, single molecule detection in medical research and diagnostics \cite{Patra_Plasmofluidic_2014}, enhancement of nanoscale light emitters \cite{Pan_Surface_2016}, wavelength-dependent light trapping used in plasmonic color filters and display pixels \cite{Goh_Threedimensional_2014, Tan_Plasmonic_2014}, and real-time monitoring of molecular binding dynamics \cite{Homola_Surface_1999} highlight its broad applicability.
While promising, these diverse applications are limited by the cost and complexity of optical detection schemes required to monitor plasmonic responses.
Importantly, they also all face the challenge of resistive losses that are inherent in any free-electron based system  \cite{khurgin_how_2015, Khurgin_Scaling_2011}.
This is a fundamental limitation in applications that require highly efficient plasmonic energy conversion, such as plasmonic enhancement of nonlinear optical phenomena or energy harvesting \cite{baffou_mapping_2010}.
Recently however, these thermal losses and the associated localized heating have attracted attention as a useful consequence of surface plasmons in metals \cite{ndukaife_plasmonicsturning_2016, Jones_Photothermal_2018}.
This includes inducing cell death in photothermal cancer therapy \cite{Huang_Plasmonic_2008, chen_targeting_2013}, initiating well controlled chemical reactions by increasing the local thermal energy density \cite{Christopher_Visiblelightenhanced_2011,baffou_nanoplasmonics_2014}, and evaporating water for purification and desalination purposes \cite{Zhou_3D_2016}.
\begin{figure}[!h]
    \centering
    \includegraphics[width=.2\textwidth]{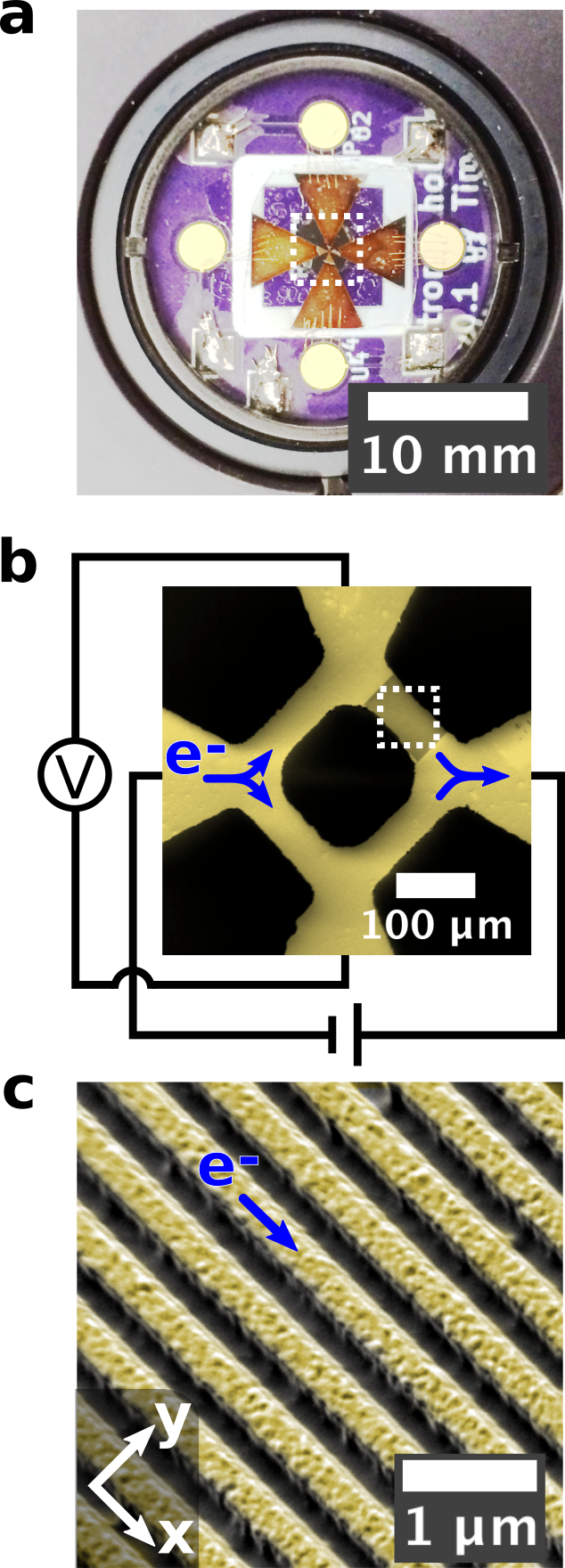}
    \caption{Images of $\mu$WB nanograting under various magnifications. Each subsequent subfigure is a zoom of the white dashed square in the previous subfigure. (a) Photograph of the mounted sample with the triangular contact pads of the $\mu$WB wire bonded to a circuit board for electrical connection. (b) False color scanning electron microscope (SEM) image of the $\mu$WB. The black areas denote the glass substrate on which the gold film (light yellow regions) is deposited. The darker yellow region enclosed in the white dashed box is the plasmonically active nanograting, etched with a focused ion beam. The electrical connections are shown schematically, with a battery connected across the left and right pads, while the induced voltage drop is measured across the top and bottom pads. The direction of current flow is indicated with blue arrows. (c) False color SEM image showing the detailed structure of the gold plasmonic nanograting with a period of 520 nm (light yellow), and the coordinate system used to define the incident electric field orientations.}
    \label{fig:sample}
\end{figure}
Additionally, it has been shown that the introduction of surface plasmons can lead to a change in a bias current passed through the plasmonic metal nanostructure itself \cite{kim_enhanced_2015, herzog_thermoplasmonics:_2014}.
We show that these effects can be used for an all-electronic readout of plasmonic activity, thus greatly simplifying and reducing the cost of plasmonic systems.
This is achieved by leveraging the scattering and subsequent thermalization of plasmonically induced hot electrons within an applied bias current in a metal nanograting. 
This offers a novel approach for direct electronic readout of light properties including polarization, which is
currently only accessible using complex optical systems \cite{Garcia_Bioinspired_2017, Mehta_Dissection_2016}.
In contrast, we demonstrate direct polarization imaging in a fully-integrated CMOS compatible plasmonic detector.
Furthermore, we show that such a detector has a response time governed by the ultrafast scattering processes involved in the relaxation of the plasmonically excited hot electrons. 

\section{Polarization dependent resistance change}
\begin{figure}[h]
    \centering
    \includegraphics[width=1\textwidth]{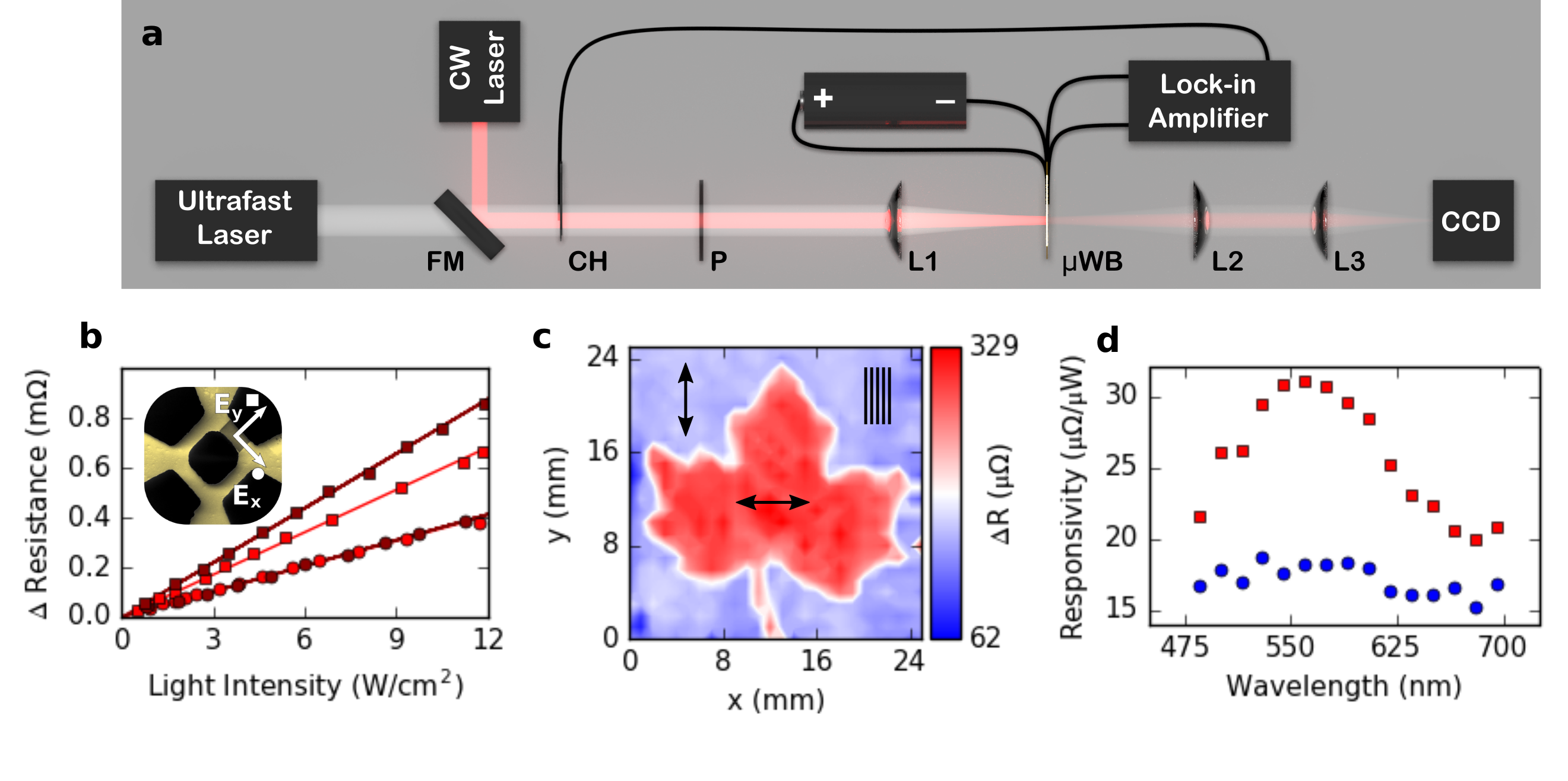}
    \caption{Optical setup and resistance data for the $\mu$WB. (a) The optical setup used to quantify the polarization and wavelength dependent resistance changes of the Au and Al $\mu$WBs. FM: flip mirror; CH: optical chopper; P: polarizing element (polarizer or mask); L1: focusing microscope objective; L2, L3: Imaging lenses. (b) Change in resistance of the Au $\mu$WB under x- and y-polarization (circles and squares, respectively) at wavelengths of 785 nm and 980 nm (light red and dark red, respectively), with the solid lines showing the linear fits. The electric field orientations with respect to the plasmonic nanograting are defined in the SEM image shown in the inset. (c) A false color polarization image of a maple leaf shaped polarization mask generated using the Au $\mu$WB under 785 nm cw illumination. Black arrows indicate the polarization direction of each region of the mask, with respect to the grating orientation shown in the upper right. The color scale corresponds to the change in resistance across the $\mu$WB. (d) Normalized resistance change of the Al $\mu$WB using a tunable ultrafast laser for x- and y- polarizations (blue circles and red squares, respectively).}
    \label{fig:Data}
\end{figure}

In order to examine the resistance change of a metal nanograting due to surface plasmon excitation, we integrate a plasmonic nanograting into a micro-Wheatstone bridge ($\mu$WB) configuration (Fig. \ref{fig:sample}). 
The optically active 100 \si{\micro\meter} x 50 \si{\micro\meter} plasmonic nanograting was fabricated on one of the four arms of a 50 nm thick Au or Al $\mu$WB on a glass substrate. 
The nanogratings have a period of 520 \si{\nano\meter} (Au) or 330 \si{\nano\meter} (Al), and can be seen as a dark-yellow area on the upper-right arm of the Au $\mu$WB shown in Fig. \ref{fig:sample}b, where the electronic connections are shown schematically.
A battery supplies 4 \si{\milli\ampere} of bias current between the left and right contacts of the $\mu$WB, while a lock-in amplifier measures an induced voltage across the top and bottom contacts that is proportional to the resistance change of the plasmonic nanograting in thermal equilibrium.
The ohmic resistance of the nanograting is on the order of 50\si{\ohm} and the materials involved are entirely CMOS compatible, making these structures amenable to fabrication as part of a larger, low impedance integrated circuit.
A description of the grating design and fabrication can be found in the methods section.
The coordinate system used throughout this work as well as the direction of net electron flow through the grating is shown in Fig. \ref{fig:sample}c. 
By measuring the voltage drop due to plasmonically induced changes in resistance, we examine the electronic signature of surface plasmons in terms of dependence on incident light intensity, polarization, and incident wavelength using the setup depicted in Fig. 2a.
A flip mirror (FM) is used to select light from either a cw diode laser or a tunable ultrafast laser (pulse lengths of 150 \si{\femto\second}) with photon energies between 1.26 and 2.55 \si{\electronvolt}, and thus much less than the work functions of Al or Au \cite{Haynes_CRC_2014}.
The beam then passes through a polarization element (P), either a polarizer for setting the beam polarization in x- or y- directions, or a polarization mask in the shape of a maple leaf for imaging purposes.
The light is then focused on the nanograting of the $\mu$WB using a microscope objective (L1) with a spot size FWHM of approximately 40 \si{\micro\meter} in order to cover the full width of the nanograting.
Figure \ref{fig:Data}b shows a comparison of resistance changes induced by orthogonal linear polarizations of cw illumination at wavelengths of 785 \si{\nano\meter} and 980 \si{\nano\meter}, incident on the Au $\mu$WB.
Regardless of polarization or wavelength, increasing the light power incident on the sample leads to a linear increase in resistance change.
Looking at the polarization dependence, the response to y- polarized light is approximately twice that of the x-polarized response. We attribute the additional polarization dependent resistance change to increased scattering of the bias current induced by the excitation of surface plasmons under y-polarized illumination.
Moreover, this additional change in resistance is dependent on the wavelength of light, with 980 \si{\nano\meter} illumination (dark red) inducing a 24\% larger resistance change than 780 \si{\nano\meter} illumination (light red).
In contrast, the resistance induced by x-polarized light is independent of wavelength, suggesting a bolometric origin \cite{Vogl_Generalized_1962}.
From the results shown in Fig. \ref{fig:Data}b, we calculate a sensitivity of 34 \si{\micro\ohm\centi\meter\squared\per\watt} and 57 \si{\micro\ohm\centi\meter\squared\per\watt} for x- and y- polarized 785 \si{\nano\meter} light, respectively.

\begin{figure}[h!]
    \centering
    \includegraphics[width=.75\textwidth]{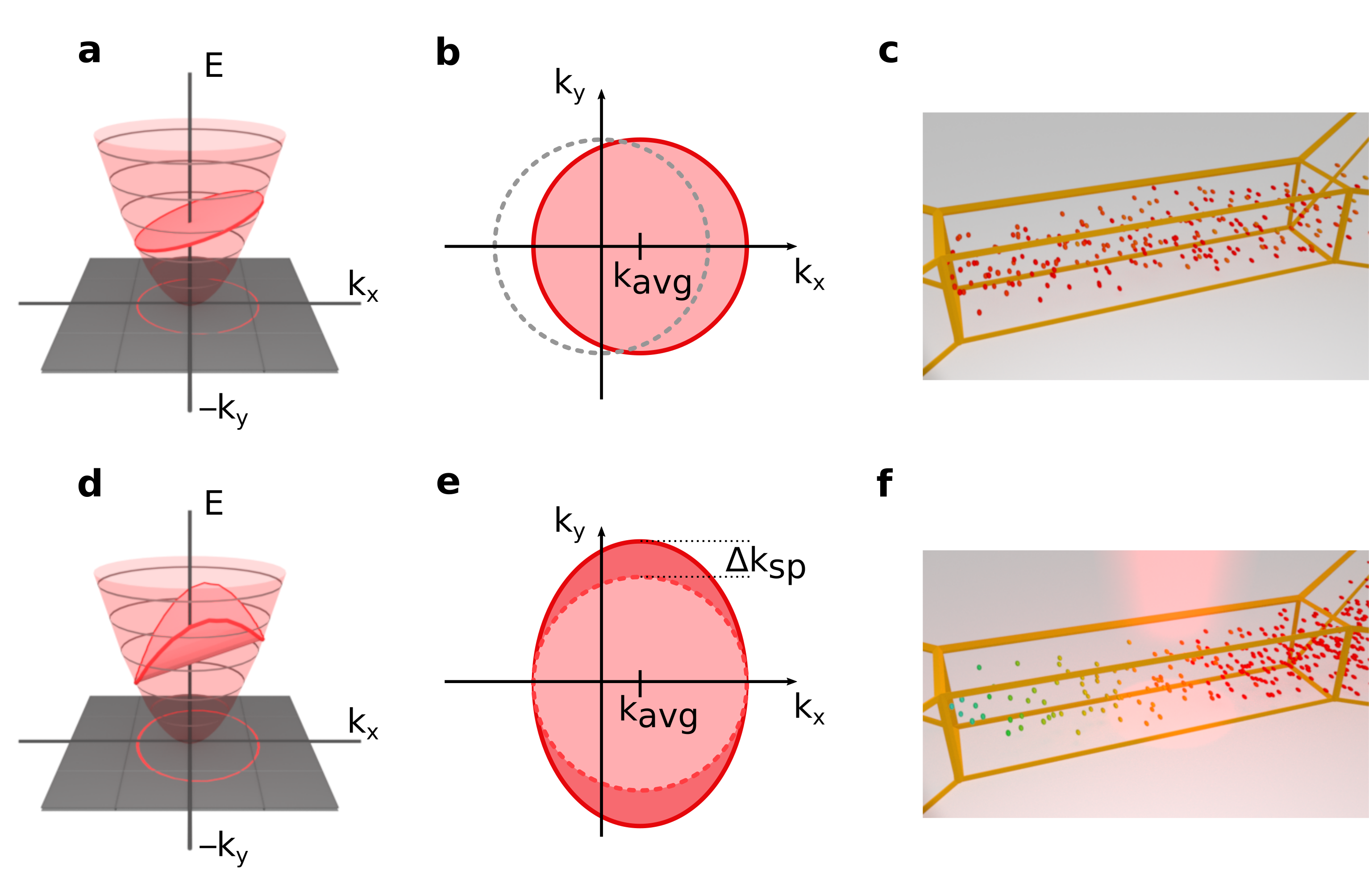}
    \caption{Physical mechanism of plasmonically induced resistance changes.(a,d) Dispersion relations of the free electron gas in $k$-space. (a) the Fermi level indicated by the oblique red surface within the energy paraboloid is tilted in the $k_\textrm{x}$-direction due to the applied bias current. (d) Plasmonically excited hot electrons lead to higher occupied states indicated by the curved wings in the $k_\textrm{y}$-direction of the red oblique surface. The 2D projection of the highest occupied state in $k$-space is shown as a red line on the grey plane and in (b,e). (b) The average momentum ($k_\textrm{avg}$)  of the free electrons in the system is shifted to the right along the x-axis by the applied bias current with respect to the equilibrium condition denoted by the grey dashed circle. (e) The plasmonic excitation of hot electrons adds an additional momentum to the free electron gas by the plasmonic wave-vector $\Delta k_{\textrm{sp}}$, oriented perpendicular to the grating. (c, f) 3D real space visualization of the electrons (spheres) of the applied bias current flowing from left to right. The color of the spheres corresponds to the x-component of the velocity where red is slow and turquoise is fast. (c) In the absence of a plasmonic disturbance, the electron flow is uniform, without a gradient in the average x-velocity component.  (f) The red illumination indicates incident laser light for plasmonic excitation. The induced increase in resistance of the system requires a larger potential drop (higher velocity) in order to maintain the same current as in (c), and leads to both a gradient of the x-velocity component as well as changes in the relative spatial distribution of the electrons.}
    \label{fig:mechanisms}
\end{figure}

The polarization dependent plasmon-induced resistance change can be understood as follows. 
Driving a small current through the $\mu$WB induces an asymmetry in the Fermi level of the free electrons in the metal grating \cite{Kasap_Principles_2005}, as shown in the band diagram in Fig. \ref{fig:mechanisms}a.
This asymmetry in Fermi level leads to a shift in the electron distribution in $k$-space in the direction of the bias current (Fig. \ref{fig:mechanisms}a,b).
At room temperature, resistance in metals is governed primarily by electron-phonon interactions and in thermal equilibrium this scattering leads to a finite resistance in metal structures.
When light is introduced into the system, photons interact directly and indirectly with the electrons in the metal, which leads to hot electron generation \cite{khurgin_how_2015}.
However, in the case of plasmonic excitation, the generation of hot electrons has a well-defined dependence on the $k$-vector of the incident light \cite{Kurosawa_Surface_2012}, forming curved wings in the Fermi level (Fig. \ref{fig:mechanisms}d)
This leads to an anisotropic, elliptical spreading of the 2D Fermi surface, shown as dark regions in  Fig. \ref{fig:mechanisms}e. 
Although the absolute value of this additional momentum $\Delta k_{\textrm{sp}}$ is on the same order of magnitude as the shift in the average momentum of the system $\Delta k_{\textrm{avg}}$ due to the applied DC bias, the major contribution of $\Delta k_{\textrm{sp}}$ is in the y-direction while the net momentum shift $k_{\textrm{avg}}$ is in the x-direction.
The introduction of $k_{\textrm{avg}}$ means that not every point within the darker regions in Fig. \ref{fig:mechanisms}e has a point with an equivalent (but negative) $k_{\textrm{x}}$ inside the light red Fermi surface, and so some negative momentum is required in the system for relaxation of the surface plasmon to occur.
This additional negative momentum originates from scattering processes within the metal and opposes the current, resulting in a measurable resistance to the applied current \cite{Kittel_Introduction_1966}.
Furthermore, as shown in the band structure in Fig. \ref{fig:mechanisms}d, the additional momentum from the generated hot electrons does not contribute equally to the total momentum of the system due to the asymmetry of the Fermi level, and the net result is a restoring of the current to the equilibrium condition.
Due to both the increase in scattering and the addition of anisotropic momentum from plasmonically generated hot electrons, we see an increase in the resistance of the system under plasmonic excitation.
Importantly, in the plasmonic case this resistance depends on the $k$-vector, and thus the wavelength and polarization, of the incident light. 
To visualize this process, Figs. \ref{fig:mechanisms}c and f show the effective speed in the x-direction of the electrons within a single ridge of the nanograting represented as color, where red is slow and turquoise is fast.

\section{Polarization imaging and wavelength sensitivity}
This polarization dependent sensitivity can be leveraged for an imaging technique utilizing polarization.
In Fig. \ref{fig:Data}c, we image a custom polarization mask in the shape of a maple leaf.
The mask consists of a 25 \si{\milli\meter} maple leaf shaped polarizing film with the transmission axis oriented in the y-direction, surrounded by a polarizing film with transmission axis oriented in the x-direction  (see supplemental material for mask fabrication details).
The laser is then raster scanned across the maple leaf mask and the induced resistance of the Au  $\mu$WB is plotted as a function of spatial coordinates (Fig. \ref{fig:Data}c).
The size of each image pixel is 1 \si{\milli\meter} in diameter, due to the FWHM of the collimated laser beam passing through the polarization mask.
The laser polarization is oriented at 45\si{\degree} to the transmission axis of the polarization mask to ensure that a constant transmission power of 80 \si{\micro\watt} is maintained for all positions. 
Using the standard deviation of this data set, we determine an upper limit for the angular resolution for linearly polarized light using a single plasmonic detector, which was found to be 4\si{\degree}.
In order to allow a full characterization of an arbitrary combination of linear polarization state and incident power, an array of 4 gratings oriented at 45 \si{\degree} to each to other could be used \cite{Qi_Real_2017}.
This is of interest where integrated polarization imaging is desirable for stability and cost/weight considerations, such as satellite-based remote sensing applications \cite{Peralta_Aerosol_2007,Hou_Improving_2018}. 

As a result of $k$-vector considerations, surface plasmons show a strong (tunable) dependence on the structural geometry of the system.
To illustrate this, Fig. \ref{fig:Data}d shows the wavelength dependent resistance change of the Al $\mu$WB under x- and y- polarized illumination using a spectrally tunable ultrafast laser.
The laser is tuned across the visible spectrum (see methods) with constant average incident light intensity, and the responsivity of the Al $\mu$WB is plotted for each wavelength.
This reveals the full character of a plasmonic resonance near 550 \si{\nano\meter}. Specifically, for y-polarized illumination (red squares) there is an increasing resistance change near the resonance, while the resistance change under x-polarized illumination (blue circles) is relatively independent of wavelength. 
By leveraging this plasmonic dependence on structure, we can directly measure the polarization and wavelength of incident light without sophisticated optical or image processing schemes.
Therefore, this electronic characterization technique offers a novel approach in applications such as hyperspectral imaging \cite{Gorbunov_Polarization_2018}, polarimetry \cite{Snik_overview_2014}, and THz detection \cite{Chen_Nanostructured_2016}. 

\section{Ultrafast dynamics}
\begin{figure}[h]
    \centering
    \includegraphics[width=.6\textwidth]{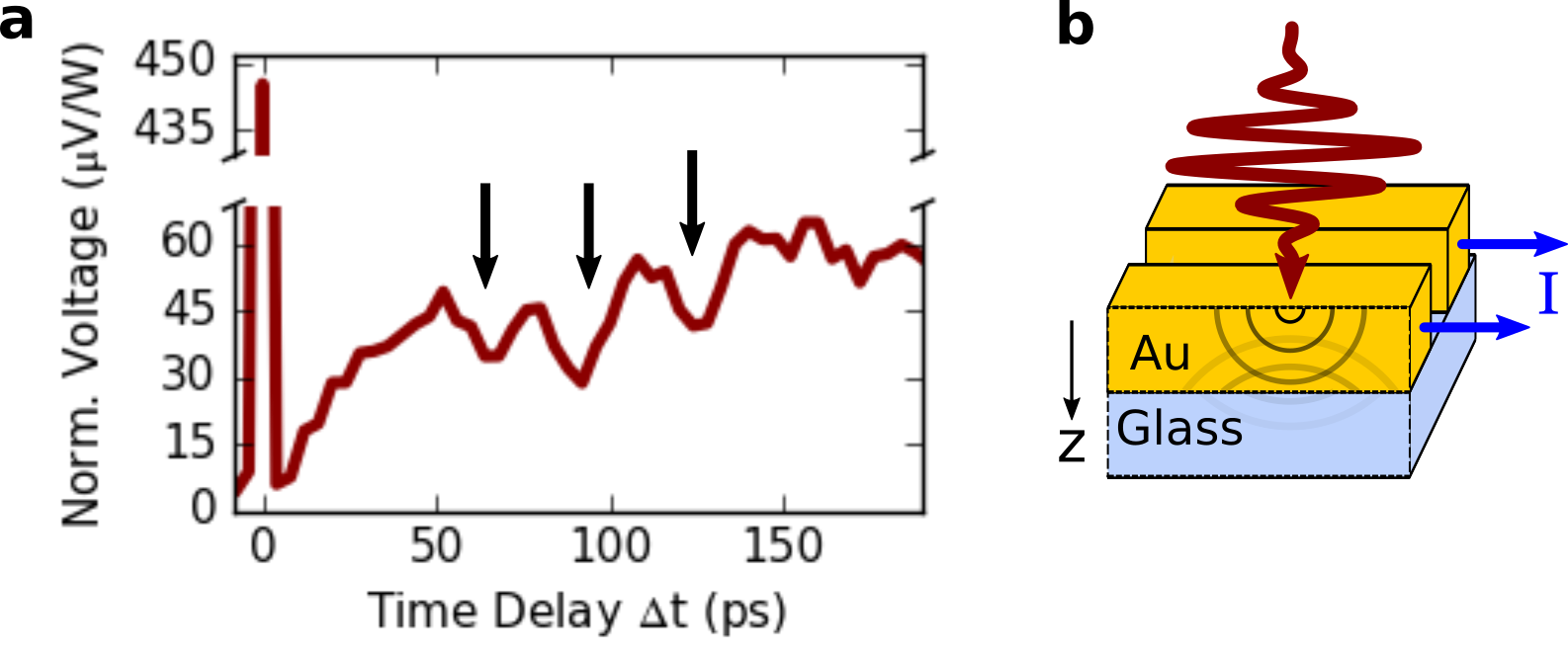}
    \caption{Time-resolved electronic readout of plasmonic hot electron interactions. (a) Power-normalized voltage drop across the nanograting induced by two consecutive ultrafast laser pulses with a time delay $\Delta t$. To accommodate the discrepancy in scale of the signal at $\Delta t=0$ without losing detail for $\Delta t>0$, the y axis is split into two regions. The arrows denote a periodic modulation of the induced voltage with a period of 30 ps, with a possible mechanism shown in (b). An acoustic shock wave induced by the first incident laser pulse propagates in the z-direction through the Au nanograting and is reflected at the Au-glass interface. The shock wave then reaches the Au-air interface at the same time as the second incident laser pulse, leading to a periodic modulation of the voltage for $\Delta t$ between 50 and 150 ps. Current direction is shown as blue arrows.}
    \label{fig:ultrafast}
\end{figure}

Timescales for plasmonically induced electron scattering rates in metals range from femtoseconds for Landau damping  \cite{Kale_Plasmons_2015}, hundreds of femtoseconds for electron-electron scattering  \cite{Zavelani-Rossi_Transient_2015}, or picoseconds for electron-phonon scattering \cite{Park_Ultrafast_2007}.
These lifetimes are typically measured optically, e.g. using T and R \cite{kolomenskii_femtosecond_2013,Harutyunyan_Anomalous_2015}, or by photoexcitation \cite{Fann_Direct_1992}, and in general only consider scattering interactions near the surface of the film within the skin depth \cite{Manke_Measurement_2013}.
However, the thermalization of hot electrons propagates throughout the Au film on the order of the Fermi velocity ($10^8$ \si{cm/s}) \cite{Brorson_Femtosecond_1987}.
Here, we examine the interaction of plasmonically induced hot electrons, and subsequently launched phonons, with a DC bias current that is distributed uniformly throughout the grating cross-section on ultrafast timescales.
In Fig. \ref{fig:ultrafast}a, we show the plasmonically induced voltage drop across the nanograting as a function of the time delay $\Delta t$ between two 800 \si{\nano\meter} light pulses with a 150 \si{\femto\second} pulse duration $\tau$, polarized in the y-direction (see supplemental information for optical setup). At $\Delta t = 0$ the two laser pulses arrive at the sample at the same time.
On timescales that are shorter than the incident laser pulses, we see a strong signal resulting from a combination of coherent processes \cite{Heinz_Coherent_1984} and electron-electron scattering that occurs during Landau damping \cite{Kale_Plasmons_2015}, where momentum is transferred between surface plasmons and hot electrons, leaving the electronic system in a highly non-equilibrium state. 
For $\Delta t > \tau$ the second laser pulse induces a scattering of the bias current that depends on $\Delta t$.
This is characterized by a transition from a shot noise regime originating from hot electron scattering \cite{Steinbach_Observation_1996} to the beginning of thermalization, where the luke-warm tail of the hot electron distribution couples to the phonons of the system \cite{brongersma_plasmon-induced_2015}.
This rise is followed by a plateau in the resistance change beginning around 20 \si{\pico\second} and extending for the remaining range of our optical delay stage.
The plateau features a modulation with a period of 30 \si{\pico\second}, indicated by arrows in Fig \ref{fig:ultrafast}a. We attribute this modulation to acoustic shockwaves induced by the ultrafast laser pulse that travel through the metal film in the z-direction, as shown in Fig \ref{fig:ultrafast}b  \cite{vanExter_Ultrashort_1988}.
When the second laser pulse arrives at the Au-air interface at the same time as a reflected propagating acoustic wavefront generated by the first laser pulse, the lattice constant and thus permittivity of the Au is altered.
This in turn affects the plasmonic response to the second pulse \cite{Wang_Resolving_2007,Temnov_Femtosecond_2009} and reduces the induced resistance of the system.
Assuming a speed of sound of $3,275$ \si{\meter\per\second} for polycrystalline gold \cite{Auld_Acoustic_1990}, a modulation of 30 ps indicates a gold thickness of 49 \si{\nano\meter} in the z direction.
This agrees with our deposited film thickness of 50 \si{\nano\meter}, as measured with a scanning electron microscope.
Therefore, this plasmonic detector shows potential as a low-impedance, CMOS compatible photodetector that has a time dependent response on the order of picoseconds for a varying optical signal, which would allow a bandwidth near THz for applications in future generations of optical communications. 

\section{Conclusion}
In summary, we have shown that changes in electronic resistance induced by plasmonic losses in a metal nanograting depend on the wavelength, polarization, and intensity of incident light.
This enables the direct detection of light properties that are only indirectly available to semiconductor detectors, offering potential applications in optical communications and imaging.
By employing different illumination schemes as well as a variety of materials and geometries, we illustrate the robustness of this technique for quantifying plasmonic responses.
Moreover, using a DC current to probe hot electron dynamics allows us to examine the interaction between excited surface plasmons and subsequent electron thermalization within the metal film in a direct way.
Since plasmonic excitation is highly sensitive to local changes in the optical properties of the surrounding medium \cite{Mock_Local_2003}, this method may offer a novel way to examine changes in refractive index that occur on fast time scales, for example in monitoring protein folding or changes in molecular conformation in real-time \cite{jiang_microsecond_2015, Hall_Conformation_2011, Stadler_Picosecond_2016}.
More generally, the electronic readout of plasmonic interaction would greatly simplify plasmon-based biosensing and chemical detection schemes by eliminating the need for highly optimized optical systems, simultaneously reducing costs and yielding more reproducible and reliable data.
Finally, because the materials, structural dimensions, and measurement tools involved are entirely compatible with current CMOS integrated circuit technology, this approach offers a means to realize truly on-chip plasmonic devices, thus enabling broader access to plasmonic technologies.

\section{Methods}
\subsection{Sample design and fabrication}
Modelling was performed using the commercial software COMSOL (FEM) and RSoft (RCWA)  to determine appropriate grating parameters for plasmon resonances in the desired spectral regions. 
Optical constants for Al and Au were taken from ref. \cite{Rakic_Optical_1998}, while the refractive index of glass was set to be 1.51. 
For sample fabrication, S1805 photoresist was spin-coated on a glass substrate, followed by a pre-bake at 115\si{\celsius} for 60 \si{\second}. 
The $\mu$WB structure was then patterned in the photoresist with UV lithography via a Karl Suss mask-aligner. 
After developing the exposed $\mu$WB pattern, Ti (3 \si{\nano\meter} adhesion layer) and then Al or Au (50 \si{\nano\meter}) was deposited using a Kurt J. Lesker E-beam evaporator 
The $\mu$WB was then protected with a drop of PMMA, and 150 nm of Al was deposited on the contact pads for better electrical contact, followed by lift-off in acetone for 12 hours.
The sample was then mounted and wire bonded to a custom printed circuit board (PCB). 
Finally, the plasmonic nanograting was milled into the $\mu$WB using a Zeiss Crossbeam focused ion beam with a current of 120 \si{\pico\ampere}.

\subsection{Electronic measurements}
The left and right contact pads of the sample (as shown in Fig. \ref{fig:sample}) were connected via a shielded cable to a battery and 1 \si{\kilo\ohm} potentiometer to set the bias current to 4 \si{\milli\ampere}. On the detection side, the top and bottom contacts were connected to a low-noise pre-amplifier. The A and B outputs of the pre-amplifier were run through twisted BNC cables to a lock-in amplifier operating in A-B configuration. The lock-in amplifier was referenced to an optical chopper operating at maximum speed (3.5 \si{\kilo\hertz}) to minimize thermal contributions to the resistance change of the nanograting.

\subsection{Ultrafast measurements} 
Wavelength dependence and ultrafast dynamics were measured using a Ti:Sapphire oscillator with a regenerative amplifier and an optical parametric amplifier (OPA). The pulse train from this system has a repetition rate of 250 \si{\kilo\hertz}. To quantify the wavelength dependence, the OPA was tuned from 486 \si{\nano\meter} to 695 \si{\nano\meter}, with a pulse fluence of 2.2 \si{\micro\joule\per\centi\meter\squared} for all wavelengths. The resistance changes were normalized to the incident power, resulting in a responsivity of the detector at each wavelength. 
The setup used for examining the ultrafast dynamics is shown in supplemental materials Fig. S2, where the pulse fluence was 28 \si{\micro\joule\per\centi\meter\squared}. The electronic readout of the ultrafast resistance dynamics is the average response of the $\mu$WB to multiple pulse pairs.

\bibliographystyle{naturemag}
\bibliography{Plastronic_manuscript_references}
\end{document}